\documentclass[a4paper]{article}
\usepackage{tikz}
\usetikzlibrary{matrix}
\usepackage{amscd,amsmath,amssymb}
\usepackage{bbm}

\newcommand{\p}[1]{(\ref{#1})}

\newcommand{\cD}{{\cal D}}
\newcommand{\cK}{{\cal K}}

\newcommand{\cQ}{{\cal Q}}

\newcommand{\cS}{{\cal S}}

\newcommand{\cP}{{\cal P}}

\newcommand{\cJ}{{\cal J}}
\newcommand{\hJ}{{\hat J}}

\newcommand{\cN}{{ {\cal N}   }}

\newcommand{\tSigma}{{\widetilde \Sigma}}

\def\hZ{\widehat{Z}}

\def\hW{\widehat{W}}

\def\hJ{\widehat{J}}
\def\hT{\widehat{T}}
\def\hV{\widehat{V}}

\newcommand{\be}{\begin{equation}}
\newcommand{\ee}{\end{equation}}
\newcommand{\bea}{\begin{eqnarray}}
\newcommand{\eea}{\end{eqnarray}}

\newcommand{\ba}{\begin{aligned}} \newcommand{\ea}{\end{aligned}}

\def\im{{\rm i}}

\newcommand{\nn}{\nonumber}

\begin{document}
\begin{flushright}
\end{flushright}\vspace{1cm}
\begin{center}
{\Large\bf New variants of $N=3,4$ superconformal mechanics}
\end{center}
\vspace{1cm}

\begin{center}
{\large\bf  Nikolay Kozyrev${}^a$ and Sergey Krivonos${}^{a,b}$}
\end{center}

\vspace{0.2cm}

\begin{center}
{ \it ${}^a$
Bogoliubov  Laboratory of Theoretical Physics, JINR,
141980 Dubna, Russia}
\vspace{0.5cm}

{\it ${}^b$ Laboratory of Applied Mathematics and Theoretical Physics, TUSUR,\\ Lenin Ave. 40, 634050 Tomsk, Russia}

\vspace{0.5cm}
{\tt nkozyrev@theor.jinr.ru, krivonos@theor.jinr.ru}
\end{center}
\vspace{2cm}

\begin{abstract}\noindent
We construct superconformal mechanics with $N=3$ and $N=4$ supersymmetries that were inspired by analogies with the supersymmetric Schwarzian mechanics. The Schwarzian, being another system with superconformal symmetry, provides insight into the field content of supersymmetric mechanics, most notably, on the number and properties of the fermionic fields involved. Adding more fermionic fields (four in the $N=3$ case and eight in the $N=4$ case) made it possible to construct systems possessing maximal superconformal symmetries in $N=3$ and $N=4$, namely $OSp(3|2)$ and $D(1,2;\alpha)$.
In the case of $N=4$ supersymmetry,  we explicitly construct a new variant of $N=4$ superconformal mechanics in which
all bosonic subalgebras of $D(1,2;\alpha)$ superalgebra have bosonic realization. In addition, the constructed systems involve $so(3)$ currents whose parametrization is not fixed, which allows to consider different underlying geometries.
\end{abstract}

\vskip 1cm
\noindent
PACS numbers: 11.30.Pb, 11.30.-j

\vskip 0.5cm

\noindent
Keywords: superconformal mechanics , $N$--extended supersymmetry, Integrable system, Schwarzians

\newpage

\setcounter{equation}{0}
\section{Introduction}
Recently, new interest in superconformal mechanics has arisen due to the intense search for $AdS_2$ solutions of both Type II and eleven dimensional supergravities.  Several recent works have classified
possible ten- and eleven-dimensional $AdS_2$ solutions with different numbers of preserved supersymmetries \cite{N3_1,N3_2,N4_1,N4_2}. These solutions are especially interesting due to
the high dimensionality of the internal space, which offers many possibilities for realizing supersymmetry.
The presence of $AdS_2$ factors leads to dual superconformal quantum mechanics with different numbers of supersymmetries.
These dual superconformal quantum mechanics should be the basis for a microscopical description of black holes with these geometries near the horizon. Unfortunately, until now mainly superconformal mechanics
with $\cN=4$ and/or $\cN=8$ supersymmetries have been fully analysed.

For example, even the simplest case of $\cN=3$ superconformal mechanics,
to our knowledge, have not been studied so far in the literature.
This forces us to construct and analyze $\cN=3$ superconformal mechanics (Section 2). Our analysis shows that one needs to deal with four fermionic components to describe such a system.
Moreover, under such circumstances, there is no possibility to avoid extending $OSp(3|2)$ supersymmetry to
$OSp(4|2)$ one.

It should be noted that two variants of $\cN=3,d=1$ superconformal
systems were constructed in the papers \cite{AG4,kk1} within the so-called
supersymmetric Schwarzian mechanics. However, a direct relation between superconformal and super Schwarzian mechanics was not established. In Section 3 we provide a detailed consideration of
the $\cN=3$ Schwarzian mechanics and demonstrate its equivalence to
the $\cN=3$ mechanics considered in Section 2. The most interesting
result from this section are the irreducibility constraints on four
$\cN=3,d=1$ superfields leading to the supermultiplet $(4,4,0)$.

Despite the fact that $\cN=4$ superconformal mechanics with
$D(1,2;\alpha)$ dynamical symmetry were widely explored in
the literature [7-22], in all the cases only one $su(2)$ subalgebra
from the $sl_2 \times su(2)\times su(2)$ bosonic core of the $D(1,2;\alpha)$
superalgebra has a bosonic realization. The second $su(2)$ subalgebra has only a fermionic realization. However, within the $AdS_2$
solutions of ten-dimensional supergravity with bosonic
$sl_2 \times S^3 \times S^3$ metrics both $S^3$ spheres have a bosonic
realization \cite{N3_1,N3_2,N4_1,N4_2}. In Section 4 we explicitly
construct a possible variant of dual superconformal mechanics in which
all bosonic subalgebras of the $D(1,2;\alpha)$ superalgebra have a bosonic realization. Finally, we conclude with a short review of the obtained results and with possible future developments.

\setcounter{equation}{0}
\section{$\cN=3$ superconformal mechanics}
\subsection{Basic ingredients}
To construct $\cN=3$ superconformal mechanics with $OSp(3|2)$ dynamical supersymmetry, let us introduce
dilaton $r$, corresponding momentum $p_r$ and three currents $J_i$,
which obey the following non-zero Poisson brackets:
\be\label{PB}
 \big\{ p_r,\, r \big\} =1\, ,\quad  \left\{J_i,\,J_j\right\} = \epsilon_{ijk}\, J_k .
\ee
The fields $r, p_r$ will be used to construct the $sl(2,\mathbb{R})$ part
of the $osp(3|2)$ superalgebra with the generators $H, D, K$ defined as
\bea\label{sl2}
H=\frac{1}{2} p_r^2 +\frac{\cal A}{r^2},\quad D= \frac{1}{2} r\, p_r,\quad K = \frac{r^2}{2}, \nn \\
\big\{  D,H \big\} =-H, \;\; \big\{ D,K\big\} =K, \;\; \big\{ H,K \big\} =2D.
\eea
Here the angular part of the Hamiltonian $\cal A$ depends only on the bosonic $so(3)$ currents $J_i$ \p{PB}  and fermions, which will be defined shortly \cite{tigran}.

To construct three supercharges $Q_i$ entering into the $osp(3|2)$ superalgebra, one has to introduce, as we noted in the Introduction, four fermions, three fermions  $\psi_i$ and an additional single fermion $\chi$,
which obey the following non-zero Poisson brackets:
\be\label{PBf}
\quad\big\{ \psi_i, \psi_j \big\} = \im \; \delta_{ij}\,, \quad
\big\{ \chi, \chi \big\} = \im\, .
\ee
The superconformal charges of the $osp(3|2)$ superalgebra are the realized with the help of the dilaton $r$ and fermions $\psi_i$ as
\be\label{S}
S_i = r \, \psi_i , \;\quad\; \big\{  S_i, S_j \big\} = 2\im \delta_{ij}\, K.
\ee
Using four fermions $\chi, \psi_i$ one can construct the $so(4)$ currents
\be\label{fso4}
\hJ_i =\frac{\im}{2} \epsilon_{ijk} \psi_j\, \psi_k, \quad
\hW_i = \im\, \psi_i \, \chi ,
\ee
which obey the standard $so(4)$ Poisson brackets:
\be\label{so4PB}
\left\{\hJ_i,\,\hJ_j\right\} = \epsilon_{ijk}\, \hJ_k,\quad
\left\{\hW_i,\,\hJ_j\right\} = \epsilon_{ijk}\, \hW_k,\; \quad
\left\{\hW_i,\,\hW_j\right\} = \epsilon_{ijk}\, \hJ_k.
\ee

\subsection{Supercharges and Hamiltonian}
From the previous Section we learned that in our construction of superconformal $osp(3|2)$ dynamical supersymmetry the conformal part,
i.e. the generators of the dilatation $D$, conformal boost $K$, conformal supercharges $S$ and the generators of $so(3)$ R-symmetry are already defined in \p{sl2}, \p{S}, \p{PB}, \p{fso4}. Moreover, the structure of the supercharges $Q_i$ is partially fixed to be
\be\label{Qstep1}
Q_i = p_r \, \psi_i + \frac{1}{r}\left[ (R-symmetry\; generators) \times fermions \right].
\ee
This structure of the supercharges has been advocated in \cite{kn1} and then
successfully applied to $\cN=8$ superconformal mechanics in \cite{kn2,kn3}.

Using the Anzatz \p{Qstep1}, the supercharges of $\cN=3$ superconformal mechanics can be easily found to be
\be\label{Qstep2}
Q_i = p_r\, \psi_i+ \frac{1}{r}\left( \epsilon_{ijk}\, J_j\, \psi_k + J_i \,\chi\right)\quad \rightarrow \quad \left\{Q_i, Q_j\right\} = 2\,\im\,\delta_{ij} H
\ee
where the Hamiltonian $H$ reads
\be\label{HN3}
H= \frac{1}{2} p_r^2 + \frac{1}{2 r^2} J_i \left( J_i -2 \hJ_i +2 \hW_i\right).
\ee
Note that supercharges \p{Qstep2} satisfy supersymmetry algebra without placing any algebraic constraints on the generators $J_i$. It is important that without the bosonic current $J_i$ one can construct only free supercharges
$Q_i = p_r\, \psi_i$ and a free purely bosonic Hamiltonian $H=\frac{1}{2} p_r^2 $.

To visualize  the  dynamical symmetry of the system, one has to calculate the Poisson brackets between
the Poincar\'e \p{Qstep2} and conformal supersymmetry generators $S_i$ \p{S}:
\be\label{osp}
\left\{S_i, Q_j\right\} = 2 \im \, \delta_{ij} D + \im \, \epsilon_{ijk} \left( J_k+\hJ_k \right).
\ee
Having in mind the brackets
\be
\left\{  J_i+\hJ_i, Q_j \right\} = \epsilon_{ijk} Q_k ,\quad
\left\{  J_i+\hJ_i, J_j+\hJ_j \right\} = \epsilon_{ijk}\left( J_k+\hJ_k\right),
\ee
we conclude that the generators
$\left\{ Q_i\; \p{Qstep2}, H\; \p{HN3}, S_i \; \p{S}, D,K\; \p{sl2}\right\}$  and $\left\{  J_i+\hJ_i \; \p{PB},\p{fso4}\right\}$ form the superalgebra $osp(3|2)$.

To complete this Section let us make several comments:
\begin{enumerate}
\item The unavoidable presence of four fermions in the game raises
the question of existence of the fourth supercharge extending the dynamical supersymmetry $osp(3|2)$ to the $osp(4|2)$ one. Indeed, one can immediately suggest that an additional superconformal charge is
\be
s = r\, \chi, \;\; \big\{ s,s \big\}=2\im K.
\ee
Then $q=\big\{  H,s\big\}$ appears to be the fourth supercharge:
\be\label{qosp42}
q = p_r\,\chi - \frac{1}{r} J_i \psi_i \quad \rightarrow \quad
\left\{ q, q\right\} = 2\,\im H, \; \left\{ q, Q_i\right\} =0.
\ee
\item The superalgebra $osp(4|2)$ contains an additional superconformal generator of $R$-symmetry $\hW_i$ \p{fso4} which appears in the brackets
of the generators $Q_i$ and $q$ with $s$:
\bea
\left\{s, Q_i\right\} = - \left\{S_i, q \right\}=  \im \left( J_i-\hW_i\right),\quad \left\{s,q\right\} = 2\, \im\, D \nn \\
\big\{ J_i +\hJ_i, J_i -\hW_i \big\} = \epsilon_{ijk}\big( J_k -\hW_k \big), \;\; \big\{ J_i -\hW_i, J_j -\hW_j \big\} = \epsilon_{ijk}\big( J_k +\hJ_k \big).
\eea
Thus, the generators $\{Q_i,q,S_i,s, H, D,K,J_i+\hJ_i,J_i-\hW_i\}$ span the $osp(4|2)$ superalgebra. Note that the generators of $so(3)$ subalgebra and of the $so(4)/so(3)$ coset have the same bosonic core but differ in the fermions.
\item It is natural to expect the possibility of constructing a full $D(1,2;\alpha)$ superalgebra with given fields. The corresponding supercharges read:
\bea\label{Dalpha}
&& Q_i =  p_r\,\psi_i -\frac{1}{r}\left[2\alpha \epsilon_{ijk}J_j \psi_k+
2\alpha J_i \, \chi+(1+2\alpha) \hJ_i \chi \right], \nn \\
&& q = p_r \chi +\frac{1}{r} \left[ 2 \alpha J_i \psi_i  + \frac{1+2\alpha}{3} \hJ_i \psi_i\right] , \quad
 \left\{Q_i, Q_j\right\} = 2\,\im\,\delta_{ij} H, \quad
\left\{q, q \right\} = 2\,\im\, H
\eea
with the Hamiltonian
\be\label{HDalpha}
H= \frac{1}{2} p_r^2 + \frac{1}{r^2}\left[2 \alpha^2  J_i J_i +2 \alpha J_i \hJ_i -2 \alpha J_i \hW_i -\frac{1+2\alpha}{3} \hJ_i \hW_i  \right].
\ee
Note that the realization of the generator $J_i$ in \p{Dalpha}, \p{HDalpha} is arbitrary, provided it has zero brackets with other fields involved. In particular, one can consider its realization via new bosonic fields only, new fermionic fields or combinations of both, reproducing systems constructed in \cite{FI5,FI2,FI4}.

\item Interestingly, our $\cN=3$ superconformal mechanics coincides with
the $\cN=3$ supersymmetric Schwarzian mechanics \cite{kk1} at the superfield level (see next Section).
\end{enumerate}

\setcounter{equation}0
\section{$N=3$ supersymmetric Schwarzian mechanics}
It is important to note that using the method of nonlinear realizations it is possible to construct superconformal mechanics which includes interactions with non-Abelian currents. The basic steps are very similar to  the construction of the supersymmetric Schwarzians mechanics \cite{AG4,kk1}.

Let us consider the $N=3$ case. The starting point here is the element of the $OSp(3|2)$ supergroup:
\be\label{N3g}
g=e^{\im t \cP}\, e^{ \theta_i \cQ_i }\, e^{ \lambda_j \cS_j} e^{\im z \cK} e^{i u \cD} e^{\im \phi_i  \cJ_i}.
\ee
The generators here obey the relations \cite{Dict,VP}:
\bea\label{N3SCA}
&& \im \left[ \cD, \cP \right]= \cP, \quad  \im \left[ \cD, \cK \right]= - \cK, \quad  \im \left[ \cK, \cP\right]=2 \cD,
\nn \\
&& \left\{ \cQ_i, \cQ_j \right\}=2 \delta_{ij} \cP, \quad \left\{ \cS_i, \cS_j \right\}=2 \delta_{ij} \cK, \quad
\left\{ \cQ_i, \cS_j \right\}= - 2 \delta_{ij} \cD -\epsilon_{ijk} \cJ_k,
\nn \\
&& \im \left[ \cD, \cQ_i \right] = \frac{1}{2} \cQ_i,\; \im \left[\cD, \cS_i \right] = -\frac{1}{2} \cS_i,\quad \im \left[ \cK,\cQ_i \right] = - \cS_i, \;
\im \left[\cP, \cS_i \right]= \cQ_i, \nn \\
&& \im \left[ \cJ_i, \cQ_j \right] = \epsilon_{ijk} \cQ_k, \quad \im \left[ \cJ_i, \cS_j \right] = \epsilon_{ijk} \cS_k, \quad \im \left[ \cJ_i, \cJ_j \right] = \epsilon_{ijk} \cJ_k.
\eea
Here, $t$, $\theta_i$ are the coordinates of the $N=3$, $d=1$ superspace, and $u$, $z$, $\phi_i$, $\lambda_i$ are supposed to be superfields on this space.

The Cartan forms invariant with respect to left multiplication are defined as
\be\label{N3cfdef}
\Omega = g^{-1}d g = \im \omega_P \cP+ \left(\omega_Q\right)_i \cQ_i + \im\omega_D \cD+
 \im \left(\omega_J\right)_{i} \cJ_{i}   +  \left(\omega_S\right)_i \cS_i + \im \omega_K \cK
\ee
and explicitly read
\bea\label{N3cf}
\omega_P &=& e^{-u} \left( dt + \im\, d\theta_i \, \theta_i \right)\equiv  e^{-u} \triangle t, \nn \\
\omega_D &=& d u - 2 z \triangle t - 2 \im \,d \theta_i \lambda_i , \\
\omega_K &=&  e^{u} \left( d z +z^2 \triangle t + \im \,d\lambda_i\, \lambda_i   + 2 \im \, z d \theta_i \,\lambda_i \right), \nn \\
\big(\omega_Q \big)_i &=& \big({\hat\omega}_Q \big)_j M_{ij}, \;\; \big(\omega_S \big)_i=\big({\hat\omega}_S \big)_j M_{ij}, \;\; \big(\omega_J \big)_i = \big({\hat\omega}_J \big)_j M_{ij} + \frac{1}{2}\epsilon_{ijk} dM_{jm} M_{km}, \nn \\
M_{ij} &=& \big(e^{\mathfrak{M}}\big)_{ij}, \;\; \mathfrak{M}_{ij} = \epsilon_{ijk}\phi_k,
\eea
where the hatted forms are
\bea\label{N3cfQSJhat}
\left(\hat\omega_Q\right)_i & = & e^{-\frac{u}{2}} \left( d \theta_i + \triangle t \lambda_i\right), \nn \\
\left( \hat\omega_J\right)_{i} & =&-\im \epsilon_{ijk} \left(d\theta_j \lambda_k  + \frac{1}{2} \triangle t\; \lambda_j \lambda_k \right),  \\
\left({\hat\omega}_S\right)_i &=&  e^{\frac{u}{2}} \left( d \lambda_i - \im \lambda_i \lambda_j d\theta_j + z\left( d\theta_i + \triangle t \, \lambda_i \right)\right).\nn
\eea
Note that by construction $M_{ij}$ is an orthogonal matrix, $\big( M^{-1} \big)_{ij} = M_{ji}$, $\det M =1$.

The covariant derivatives on $(t,\theta_i)$ superspace can be defined in the standard way as
\be\label{N3der}
\partial_t = \frac{\partial}{\partial t}, \;\; D_i = \frac{\partial}{\partial \theta_i} -\im  \theta_i \frac{\partial}{\partial t},  \quad \quad \left\{ D_i, D_j\right\} = - 2 i \delta_{ij} \partial_t.
\ee

Superconformal mechanics usually contains much fewer components that are present in the $\theta$-expansions of the superfields $u$, $z$, $\phi_i$, $\lambda_i$, and one should impose additional superconformally invariant irreducibility conditions. Let us take
\be\label{N3cond}
\omega_D =0, \;\; \big( \omega_J \big)_i = \omega_P \, B_i + \big(\omega_Q \big)_i \, \Sigma.
\ee
Both of these conditions were encountered in the construction of the Schwarzian mechanics \cite{kk1}. A non-trivial point in the second condition is the restriction $\Sigma_{ij}=\delta_{ij}\Sigma $ in the most general expansion of $\big( \omega_J \big)_i = \omega_P \, B_i + \big(\omega_Q \big)_j \, \Sigma_{ij}$.
 Note that the field $\Sigma$ holds the place of the $N=3$ Schwarzian.

The $\omega_D =0$ condition expresses superfields $z$ and $\lambda_i$ in terms of $u$:
\be\label{zpsiu}
z = \frac{1}{2}\dot u, \;\; \lambda_i = -\frac{\im}{2}D_i u.
\ee

The second condition in \p{N3cond} reads
\be\label{Mcond1}
\frac{1}{2}\epsilon_{ijk}D_m M_{jn} \, M_{lm}\, M_{kn} -\im M_{ij}\, M_{lm}\, \epsilon_{jmn}\lambda_n  = \delta_{li}
e^{-\frac{u}{2}}\Sigma =\delta_{li} \tSigma .
\ee

To analyze the relation \p{Mcond1}, it is useful to introduce some parametrization for the orthogonal matrix $M_{ij}=\big( e^{\mathfrak{M}}\big)_{ij}$, $\mathfrak{M}_{ij}=\epsilon_{ijk}\phi_k$. The most convenient one is just the stereographic projection
\be\label{Mij}
M_{ij} = \frac{1- \frac{1}{4}y^2}{1+\frac{1}{4}y^2} \delta_{ij} + \frac{\epsilon_{ijk}y_k}{1+\frac{1}{4}y^2}+ \frac{1}{2}\frac{y_i y_j }{1+\frac{1}{4}y^2}, \;\; y_i = \phi_i \frac{\tan\left(\frac{1}{2}\sqrt{\phi^2}\right)}{\frac{1}{2}\sqrt{\phi^2}}, \;\; y^2 =y_i y_i, \;\; \phi^2 = \phi_i\phi_i.
\ee

Substituting \p{Mij} and taking into account \p{zpsiu}, one can reduce \p{Mcond1} just to
\bea\label{ycond1}
\frac{D_j y_i + \frac{1}{2}\epsilon_{imn}y_m D_j y_n }{1+\frac{1}{4}y^2}= \delta_{ij}\tSigma + \frac{1}{2}\epsilon_{ijk}D_k u \;\;\mbox{or, equivalently,} \;\;\nn \\
D_j y_k = N_{km}\Lambda_{mj}, \;\; \Lambda_{mj}= \delta_{mj} \tSigma + \frac{1}{2}\epsilon_{mjn}D_n u , \;\; N_{ij} = \delta_{ij} + \frac{1}{2}\epsilon_{ijk}y_k +\frac{1}{4}y_i y_j.
\eea
Therefore, one can constrain the symmetric part of $D_{i}y_j$, at the same time expressing $D_i u$ in terms of the derivatives of $y_k$:
\bea\label{ircon44}
&\epsilon_{kij} \frac{D_j y_i + \frac{1}{2}\epsilon_{imn}y_m D_j y_n }{1+\frac{1}{4}y^2} =D_k u,& \nn \\
&D_i y_j +D_j y_i + \frac{1}{2}\epsilon_{imn}y_m D_j y_n  +  \frac{1}{2}\epsilon_{jmn}y_m D_i y_n = \delta_{ij}\left( \frac{2}{3} D_m y_m -\frac{1}{3} \epsilon_{mnp}y_m \, D_n y_p    \right). &
\eea
Relations \p{ircon44} are the irreducibility condition of the multiplet $(4,4,0)$ with the physical bosons being the first components of $u$ and $y_i$, while 4 fermions being $D_i u$ and $D_m y_m$ \footnote{In what follows, it is slightly preferable to use $\tSigma$ as the fourth fermion instead of $D_m y_m$.}.
To prove this statement, it is sufficient to check that $D_i D_j y_k$ and
$D_i D_j u$ can be expressed in terms of the $u$, $y_i$ superfields and their derivatives.

First, calculating $D_i D_j y_k$ from \p{ycond1}, one can relate it to $D_i \tSigma$ and $\epsilon_{jpq}D_p D_q u$
\bea\label{Dycond1}
D_i D_j y_k = D_i N_{km}\Lambda_{mj} + N_{km}D_i\Lambda_{mj} = \left( \frac{1}{2}\epsilon_{kmq}N_{qp} + \frac{1}{4}y_m N_{kp}+ \frac{1}{4}y_k N_{mp}  \right)\Lambda_{pi}\Lambda_{mj}+ \nn \\+N_{km}\left( D_i \tSigma \delta_{mj} + \frac{\im}{2}\epsilon_{ijm}\dot u - \frac{1}{4}\delta_{im}\epsilon_{jpq}D_p D_q u + \frac{1}{4}\delta_{ij}\epsilon_{mpq}D_p D_q u  \right).
\eea
On the other hand, $D_i D_j y_k +D_j D_i y_k = \big\{ D_{i}, D_j \big\}y_k = -2\im \delta_{ij} \dot y_k$. Comparing this with \p{Dycond1}, one can find that
\be\label{Dycond4}
\epsilon_{ipq}D_p D_q u = -4\im \big(N^{-1}\big)_{ij}\dot y_j -2 \tSigma D_i u, \;\; D_i \tSigma = -\im \big(N^{-1}\big)_{ij}\dot y_j + \frac{1}{8}\epsilon_{ipq}D_p u\, D_q u.
\ee
Next, expression for the second derivative of $y_k$ can be found by substituting \p{Dycond4} into \p{Dycond1}.  Thus, we demonstrate that the second covariant derivatives of $u$ and $y_k$ can be expressed in terms of
physical components and their time derivatives.

To construct an superconformally invariant action, one should study explicitly the $OSp(3|2)$ transformations. As special conformal transformations can be produced as commutators of superconformal ones, it is sufficient to find transformations induced by $g_S = e^{\eta_i \cS_i}$:
\bea\label{N3Str}
\delta_S t = -\im \eta_i \theta_i t, \;\; \delta_S \theta_i = -\eta_i t -\im \eta_k \theta_k \theta_i, \;\; \delta_S u = -2\im \eta_k \theta_k, \;\; \delta_S M_{ij} = \im M_{in}\big( \eta_n \theta_j - \eta_j \theta_n \big), \nn \\
\delta_S z = \im \eta_k \lambda_k +2\im \eta_k \theta_k z, \;\; \delta_S \lambda_i = \eta_i + \im \eta_k \theta_k \lambda_i -\im \eta_k \lambda_k \theta_i -\im \eta_i \theta_k \lambda_k.
\eea
The conformally invariant measure on the superspace can be constructed if one notes that
\be\label{N3measure}
dt^\prime d^3\theta^\prime = \mbox{Ber}\frac{\partial (t^\prime, \theta^\prime)}{\partial(t,\theta)} dt d^3\theta = dt d^3\theta\big( 1 + \im \eta_i \theta_i\big),
\ee
and, therefore, $dt d^3\theta e^{\frac{u}{2}}$ is invariant. The action should involve another conformally invariant fermionic superfield; a suitable candidate is thus \footnote{It is worth noting that this action is just the $N=3$ Schwarzian action where the variables of integration were changed from ``invariant'' $\tau,\tilde{\theta}_i$ to $t,\theta_i \sim \xi_i(\tau,\tilde\theta)$. }
\be\label{N3candact}
-12 S_{N=3} = \int dt d^3 \theta e^{\frac{u}{2}} \Sigma = \int dt d^3 \theta e^{u} \tSigma.
\ee

Straightforward but lengthy calculation of the integral over $\theta_i$ using \p{ycond1}, \p{Dycond4} results in
\be\label{N3compact1}
S_{N=3} = \frac{1}{2}\int dt e^u \left( \frac{1}{4}\dot{u}^2 + g_{ij} \dot{y}_i \dot{y}_j +\im \lambda_i \dot{\lambda}_i - \im \tSigma \dot{\tSigma} +\big(-\im \epsilon_{ijk}\lambda_j \lambda_k +2 \tSigma \lambda_i\big)\big( N^{-1}  \big)_{im}\dot{y}_m -\im \tSigma \epsilon_{ijk}\lambda_i \lambda_j \lambda_k     \right), \ee
where we used the same notation for superfields and their first components. The metric in the internal sector is
\be\label{metric}
g_{ij} = \big( N^{-1} \big)_{ki}\big( N^{-1} \big)_{kj} = \frac{\delta_{ij}}{1+ \frac{1}{4}y^2}- \frac{1}{4} \frac{y_i y_j}{\left( 1+ \frac{1}{4}y^2  \right)^2}.
\ee
Due to $SO(3)$ symmetry, one can expect the metric to describe a $3$-sphere. This becomes clear after field redefinition:
\bea\label{xchisigma}
y_i = \frac{z_i}{1- \frac{1}{16}z^2}, \;\; u= 2\log r, \;\; \lambda_i = \frac{1}{r}\psi_i, \;\; \tSigma = -\frac{\im}{r}\chi,  \;\; {\tilde N}_{ij} = \big( 1-\frac{1}{16}z^2\big) \delta_{ij} +\frac{1}{8}z_i z_j + \frac{1}{2}\epsilon_{ijm}z_m \;\; \Rightarrow  \nn \\
S_{N=3} = \frac{1}{2}\int dt\left( \dot{r}^2 + r^2 \frac{{\dot z}_i {\dot z}_i}{\big( 1 + \frac{1}{16}z^2   \big)^2} -\im \dot{\psi}_i \psi_i - \im {\dot \chi} \chi -\im \big( \epsilon_{ijk}\psi_j\psi_k +2  \chi \psi_i \big)  \big( \tilde N^{-1}  \big)_{im}\dot{z}_m  - r^{-2}\chi \epsilon_{ijk}\psi_i \psi_j \psi_k     \right).
\eea
The respective Hamiltonian, momenta and Dirac brackets read
\bea\label{Hpbr}
H = \frac{p_r^2}{2} + \frac{1}{2r^2} \frac{p_i p_i}{\left( 1+\frac{1}{16} z^2\right)^2} + \frac{\im }{2r^2}p_i {\tilde N}_{ij} \big( 2\chi \psi_j+ \epsilon_{jmn}\psi_m \psi_n   \big), \nn \\
p_r = \dot{r}, p_i = r^2 \left(1+\frac{1}{16}z^2  \right)^{-2} {\dot z}_i -\im \chi \psi_k \big({\tilde N}^{-1} \big)_{ki} -\frac{\im}{2} \epsilon_{mnp}\psi_n \psi_p \big({\tilde N}^{-1} \big)_{mi}, \nn \\
\big\{ p_r , r  \big\} =1, \;\; \big\{ p_i , z_j  \big\} = \delta_{ij}, \;\; \big\{ \psi_i , \psi_j  \big\} = \im \delta_{ij}, \;\; \big\{ \chi , \chi  \big\} = \im.
\eea
The transformation laws of the components are remarkably simple:
\bea\label{N3susytr}
\delta f|_{\theta\rightarrow 0} = (\epsilon_i D_i f)|_{\theta\rightarrow 0} \Rightarrow \;\;\delta r = \im \epsilon_i \psi_i, \;\; \delta \psi_i = -\epsilon_i p_r  -\frac{\epsilon_j \epsilon_{jik}{\tilde N}_{nk}p_n}{r}, \nn \\
\delta z_i = \frac{\im \epsilon_j }{r}\big( -\chi {\tilde N}_{ij} + {\tilde N}_{ik}\epsilon_{kjm}\psi_m  \big), \;\; \delta \chi = \frac{1}{r}\epsilon_i {\tilde N}_{ji}p_j.
\eea
The transformation laws of the fermions \p{N3susytr} contain $p_i$ and $z_i$ only as a part of combination $J_i = - {\tilde N}_{ji}p_j$. As can be checked, these currents form the $so(3)$ algebra:
\be\label{Jbracs}
\big\{ J_i, J_j \big\} = \epsilon_{ijk}J_k.
\ee

The transformations \p{N3susytr} can be reproduced via the Dirac bracket:
\be\label{N3susytr2}
\delta f = \im \big\{ \epsilon_i Q_i , f \big\}, \;\; Q_i = p_r \psi_i +\frac{1}{r}J_i \chi + \frac{1}{r}\epsilon_{ijk}J_j\psi_k.
\ee
These supercharges coincide with ones constructed in Section 2 for a particular choice of $J_i$.

\setcounter{equation}0
\section{New variant of $\cN=4$ superconformal mechanics}
Among the constructed superconformal mechanics, the different variants with $\cN=4$ supersymmetry have definitely attracted the most attention and are the most studied to date [7-22].

The most general $\cN = 4, d = 1$ superconformal group is the exceptional supergroup $D(1, 2; \alpha)$ \cite{Dict,VP}.  The realizations of
$D(1,2; \alpha)$ in the models of supersymmetric mechanics were a subject of many works (see, e.g., \cite{FI5} and references therein).
However, most of  the realizations were based on one or another fixed
type of the irreducible $\cN = 4, d = 1$ supermultiplet. Only recently, the study of superconformal systems including some pairs of such multiplets was initiated in \cite{AG1} and then continued in
\cite{FI2}. However, even in these generalizations only one $su(2)$ subgroup of $R$-symmetry has a bosonic realization. Second $su(2)$ algebra was realized purely on fermions.
In this section we construct supercharges and Hamiltonian of
$\cN=4$ superconformal mechanics with both $su(2)$ subgroups having the bosonic core.
\subsection{Basic ingredients}
To construct $\cN=4$ superconformal mechanics with $D(1,2;\alpha)$ dynamical supersymmetry, let us introduce the following set of bosonic  and fermionic fields:

\begin{itemize}
	\item  the bosonic field $r$ and the corresponding momenta $p_r$,
	to realize the conformal $sl(2,\mathtt{R})$ symmetry as in \p{sl2}
	\item two triplets of the bosonic currents $J^{ab}$ and $T^{ij}$
	obeying the Poisson brackets:
\be\label{currents}
\left\{J^{ab},\,J^{cd}\right\} = -\left(\epsilon^{ac}J^{bd} +\epsilon^{bd}J^{ac} \right), \quad
	\left\{T^{ij},\,T^{kl}\right\} = -\left(\epsilon^{ik}T^{jl} +
\epsilon^{jl}T^{ik} \right).
\ee
We do not fix the parameterization of these $su(2)$ subalgebras.
\item eight fermionic fields $\psi^{ia}$ and $\sigma^{ia}$ obeying the brackets
\be
\left\{\psi^{ia}, \psi^{jb}\right\}= 2 \im \, \epsilon^{ij} \epsilon^{ab},\quad \left\{\sigma^{ia}, \sigma^{jb}\right\}= 2 \im \, \epsilon^{ij} \epsilon^{ab}.
\ee
From these fermions one can construct four $su(2)$  algebras with the generators
\be
\hJ_\psi^{ab} = \frac{\im}{4}\psi^{i a}\psi_{i}^b, \quad
\hJ_\sigma^{ab} = \frac{\im}{4}\sigma^{i a}\sigma_{i}^b, \quad
\hT_\psi^{ij} = \frac{\im}{4}\psi^{i a}\psi^{j}_a, \quad
\hT_\sigma^{ij} = \frac{\im}{4}\sigma^{i a}\sigma^{j}_a .
\ee
All these currents obey the same brackets as in \p{currents}.
\item From the fermionic fields  $\psi^{ia}$ and $\sigma^{ia}$ one may construct additional currents
\be
\hV^{ab} =\frac{\im}{8}\left(\psi^{i a}\sigma_{i}^b+\psi^{i b}\sigma_{i}^a\right), \; \hW^{ij} = \frac{\im}{8}\left(\psi^{i a}\sigma^{j}_a+
\psi^{j a}\sigma^{i}_a\right),
\ee
which span two $so(4)$ algebras together with the generators $\hJ_\psi^{ab}+\hJ_\sigma^{ab}$ and $\hT_\psi^{ij}+\hT_\sigma^{ij}$, respectively:
\bea\label{currents1}
&&\left\{\hV^{ab},\,\hV^{cd}\right\} = -\frac{1}{4}\left(\epsilon^{ac}(\hJ_\psi^{bd}+\hJ_\sigma^{bd}) +\epsilon^{bd}(\hJ_\psi^{ac}+\hJ_\sigma^{ac}) \right), \nn\\
&&	\left\{\hW^{ij},\,\hW^{kl}\right\} = -\frac{1}{4}\left(\epsilon^{ik}(\hT_\psi^{jl}+\hT_\sigma^{jl}) +
\epsilon^{jl}(\hT_\psi^{ik}+\hT_\sigma^{ik}) \right).
\eea
\item Finally, we have the $U(1)$ current
\be
\hZ = \frac{i}{4} \psi^{ia} \sigma_{ia} .
\ee
\end{itemize}

Coming back to a possible superfield description of our set of fields note
that the fields $\{ r, \psi^{ia}, J_{ab}\}$ fit into the $\cN=4$ superfield
$(1,4,3)$. The rest of the fields form, probably, the spin 1 superfield
$(3,4,1)$ $\{T^{ij},{\dot\sigma}{}^{ia},{\cal A}\}$ with $\cal A$ being auxiliary component which is invisible within the Hamiltonian description.
All together, these components fit well into the unrestricted $N=4$ superfield $r$.

\subsection{Supercharges and Hamiltonian}
Using the Anzatz \p{Qstep1}, the supercharges of $\cN=4$ superconformal mechanics can be  found to be
\bea\label{N4Q}
Q^{i a} & = & p_r\, \psi^{i a}- \frac{1}{r}\,2\alpha\,\left( T^{ij}+\frac{1}{3} \hT_\psi^{ij}+  \hT_\sigma^{ij}\right)\psi_j{}^a+
\frac{1}{r}2\alpha \left( T^{ij}+\frac{2}{3}\hT_\sigma^{ij}\right) \sigma_j{}^a + \nn \\
&& \frac{1}{r}2 (1+\alpha) \left(  J^{ab}+\frac{1}{3}\hJ_\psi^{ab}+\hJ_\sigma^{ab}\right)\psi^i{}_b +
\frac{1}{r} 2(1+\alpha)\left(  J^{ab}+\frac{2}{3} \hJ_\sigma^{ab}\right) \sigma^i{}_b +\frac{M}{r} \sigma^{ia} .
\eea
They obey the standard brackets
\be
\left\{Q^{ia}, Q^{jb}\right\} = 4\, \im\, \epsilon^{ij} \epsilon^{ab}H
\ee
where the Hamiltonian $H$ reads
\bea\label{N4H}
H&=& \frac{1}{2} p_r^2 + \frac{1}{r^2}\left[ 2(1+\alpha)^2 J^{ab}J_{ab} +2\alpha^2 T^{ij}T_{ij} -2 (1+\alpha) J^{ab}\left((\hJ_\psi)_{ab}-(\hJ_{\sigma}){}_{ab}+2 \hV_{ab}\right)\right]+\nn \\
&& \frac{1}{r^2}\left[ 2\alpha T^{ij}\left((\hT_\psi)_{ij}- (\hT_\sigma)_{ij}-2 \hW_{ij}\right)-
\frac{1+\alpha}{3} \hJ_\psi^{ab}\left((\hJ_\psi)_{ab}+6 (\hJ_\sigma)_{ab}\right) +\frac{\alpha}{3}\hT_\psi^{ij}\left((\hT_\psi)_{ij} +6(\hT_\sigma)_{ij}\right)
\right]+\nn \\
&& \frac{1}{r^2} \left[ (1+\alpha) \hJ_{\sigma}^{ab}\left( (\hJ_\sigma)_{ab}-\frac{8}{3} \hV_{ab}\right)-\alpha \hT_{\sigma}^{ij}\left((\hT_{\sigma})_{ij}+\frac{8}{3} \hW_{ij}\right)+	2 M \hZ +\frac{M^2}{2} \right].
\eea
To visualize  the  dynamical symmetry of the system, one has to calculate the Poisson brackets between
the Poincar\'e \p{N4Q} and conformal supersymmetry generators $S^{i a}$ :
\be\label{N4S}
S^{i a} = r \psi^{i a} .
\ee
\be\label{osp42}
\left\{S^{i a}, Q^{j b} \right\} = 4 \im \, \epsilon^{ij} \epsilon^{ab} D +4\, \im \alpha \epsilon^{ab} \left( T^{ij}+\hT_\psi^{ij}+\hT_{\sigma}^{ij}\right) - 4\,\im \,(1+\alpha) \epsilon^{ij} \left( J^{ab}+\hJ_\psi^{ab}+\hJ_\sigma^{ij} \right).
\ee
Thus, we conclude that our system possesses $D(1,2;\alpha)$ dynamical symmetry. The full $R$-symmetry is $sl_2\times S^3\times S^3 $.

Let us also note that one can remove bosonic currents either $J^{ab}$ or $T^{ij}$ (or both of them) from the supercharges \p{N4Q} and check that the $D(1,2;\alpha)$ algebra is still closed. The resulting systems will, however, contain $8$ fermions, and cubic terms with the $\sigma^{ia}$ will be present in the supercharge, ensuring that $\sigma^{ia}$ can not be absorbed into the remaining bosonic current. Therefore, this way one obtains a different realization of $D(1,2,\alpha)$ compared to \p{Dalpha} and \cite{FI5,FI2,FI4}.

\section{Conclusion}
In this article we have considered the construction of superconformal mechanical systems with $N=3$ and $N=4$ supersymmetries, that were inspired by analogies with the supersymmetric Schwarzian mechanics \cite{kk1,kk2}. The Schwarzian, being another system with superconformal symmetry, provides insight into the field content of supersymmetric mechanics, most notably, into the number and properties of the fermionic fields involved. Adding more fermionic fields (four in the $N=3$ case and eight in the $N=4$ case) made it possible to construct systems possessing maximal superconformal symmetries in $N=3$ and $N=4$, namely $OSp(3|2)$ and $D(1,2;\alpha)$. In addition, the constructed systems involve $so(3)$ currents whose parametrization is not fixed, which allows one to consider different underlying geometries. The construction of the superfield $N=3$ action is also very similar to the Schwarzian one, including the group element, superfield constraints and the superfield Lagrangian.

Our study opens up possibilities of constructing new
superconformal mechanical systems, especially ones with extended
$\cN=5,6$ supersymmetries, which could be analogues of the generalized Schwarzians \cite{kk3}.

\section*{Acknowledgements} SK acknowledges partial financial support of the Ministry of Science and Higher Education of Russia,
Government Order for 2023-2025, Project No. FEWM-2023-0015 (TUSUR).

\end{document}